\documentclass[aps,amsmath,amssymb,showpacs,floatfix,twocolumn,prl,superscriptaddress]{revtex4}
\usepackage{graphicx}
\usepackage{dcolumn}
\usepackage{bm}

\def\SF{\mathrm{SF}}
\def\DMFT{\mathrm{DMFT}}
\def\pdag{{\phantom{\dagger}}}

\graphicspath{{figures/}}

\begin{document}

\title{Polarized superfluidity in the attractive Hubbard model with population imbalance}

\date{\today}

\author{Tung-Lam Dao}
\affiliation{Centre de Physique Th{\'e}orique, CNRS, Ecole
Polytechnique, F-91128 Palaiseau Cedex, France.}
\author{Michel Ferrero}
\affiliation{Centre de Physique Th{\'e}orique, CNRS, Ecole
Polytechnique, F-91128 Palaiseau Cedex, France.}
\author{Antoine Georges}
\affiliation{Centre de Physique Th{\'e}orique, CNRS, Ecole
Polytechnique, F-91128 Palaiseau Cedex, France.}
\author{Massimo Capone}
\affiliation{SMC, CNR-INFM and Universit{\`a} di Roma ``La Sapienza'', Piazzale Aldo
Moro 2, I-00185 Roma, Italy}
\author{Olivier Parcollet}
\affiliation{Institut de Physique Th{\'e}orique, CEA, IPhT, CNRS, URA 2306,
F-91191 Gif-sur-Yvette, France}

\begin{abstract}
We study a two-component Fermi system with attractive interactions
and different populations of the two species in a  cubic lattice.
For an intermediate coupling we find a uniformly polarized
superfluid which is stable down to very low temperatures. The
momentum distribution of this phase closely resembles that of the
Sarma phase, characterized by two Fermi surfaces. This phase is
shown to be stabilized by a potential energy gain, as in a BCS
superfluid, in contrast to the unpolarized BEC which is stabilized
by kinetic energy. We present general arguments suggesting that
preformed pairs in the unpolarized superfluid favor the
stabilization of a polarized superfluid phase.
\end{abstract}

\maketitle

The study of superfluid phases is a fundamental issue in condensed
matter physics. It has received a revived interest with the
experimental realization of cold atomic systems that allow to
probe such phases with a remarkable
controllability~\cite{M.Greiner:Nature415,Bloch:2005NatPhB,Jaksch:2005AnPhy.315}.
It is for instance possible to address a large range of
interaction strengths or to control the population imbalance
between atoms in different hyperfine states.  For fermionic fluids
composed of two species, the latter parameter, which introduces a
mismatch in the Fermi surfaces, raises exciting questions about
the stability of the conventional superfluid phase and the
possible generation of more exotic ones. Indeed, in the absence of
imbalance, a weak attractive interaction between the fermionic
species stabilizes a Bardeen-Cooper-Schrieffer (BCS) ground state,
with a pairing between species of opposite momentum near their
common Fermi surface. When the interaction is strong, the fermions
pair in real space, and superfluidity is associated with the
Bose-Einstein condensation (BEC) of pairs. The BEC-BCS crossover
has been studied intensively both
experimentally~\cite{S.Jochim:Science302,PhysRevLett.91.250401,PhysRevLett.93.050401}
and
theoretically~\cite{Leggett:1980,Nozieres:JLTP:1985:1,Randeria:prl:3202:1993,keller:prl:2001}.

The situation is far less clear when a population imbalance
introduces a mismatch between the Fermi surfaces.  At small
imbalance, the species are expected to still  form a standard BCS
or BEC state.  At larger imbalance, either superfluidity
disappears in favor of a polarized normal fluid or more exotic
forms of pairing occur. One candidate is the
Fulde-Ferrell-Larkin-Ovchinnikov
state~\cite{fulde:pr:1964:1,Larkin,koponen:njp:2008,yoshida:063601}
in which Cooper pairs appear at a non-zero total momentum.  At
zero temperature, two other possible phases that exhibit both a
non-zero superfluid order parameter and a finite polarization have
been proposed: the Sarma (or breached-pair BP2)
phase~\cite{sarma:jpcs:1963:1,liu:prl:2003:1} and the BP1
phase~\cite{sheehy:060401,son:013614,pilati:030401}. At
weak-coupling, the Sarma phase is unstable unless specific types
of interactions are considered~\cite{forbes:017001}. The BP1 has
been proposed as a stable ground state deep in the BEC regime of
trapped fermionic gases, where the system is described by a
Bose-Fermi mixture.  While both of these phases are polarized
superfluids with gapless excitations, their nature is different:
the Sarma phase has two Fermi surfaces while the BP1 phase has a
single Fermi surface for the unpaired fermions. These non-standard
phases are in general unstable at weak coupling, resulting in
phase separation between an unpolarized superfluid and a polarized
normal fluid formed by the excess fermions, an effect which has
been observed
experimentally~\cite{Hulet:science:2006:1,Zwierlein:science:2006:1,Zwierlein:nature:2006:1}.
At zero temperature $T=0$, the Sarma and BP1 phases are signaled
by a non-zero superfluid order parameter together with a finite
polarization. When $T>0$, this criterion is no longer valid
because a standard BCS or BEC state also acquires a small
polarization coming from thermally excited quasiparticles.

In this paper, we focus on polarized superfluid phases (pSF) in a
three-dimensional cubic lattice. We study their nature at weak and
intermediate coupling as a function of the temperature, treating
the effect of correlations beyond static mean field. Our main
result is that, at intermediate coupling, a pSF phase can be
stabilized down to very low temperatures, with properties which
are clearly associated with the Sarma phase. The mechanism
responsible for this stabilization is the reduction of the
polarizability of the normal fluid due to the existence of
preformed pairs. We will show that this phase is profoundly
different from the unpolarized BEC superfluid which holds at the
same coupling strength in the absence of imbalance.

\begin{figure}[ht!]
  \begin{center}
    \includegraphics[width=4.25cm,clip=true]{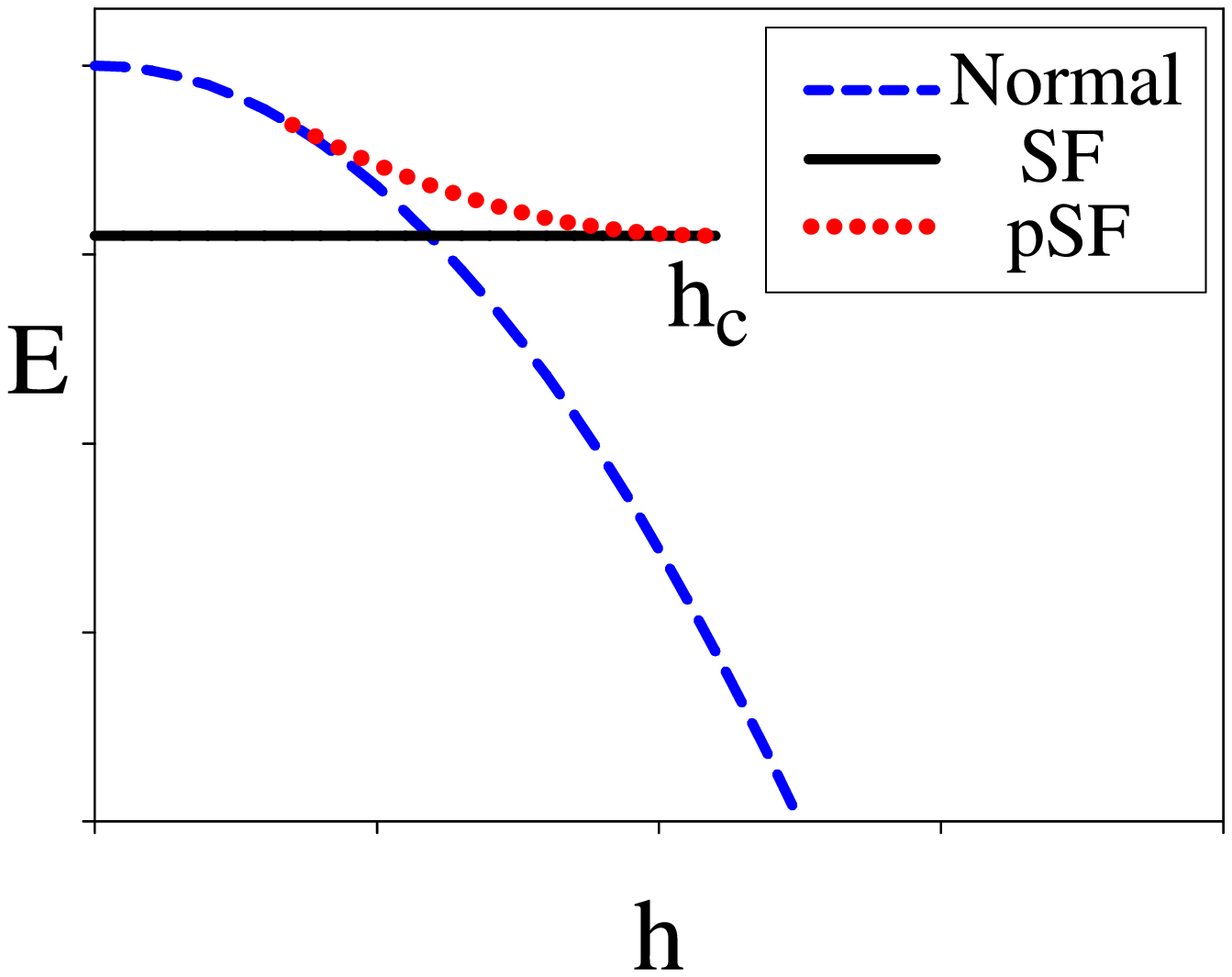}
    \includegraphics[width=4.25cm,clip=true]{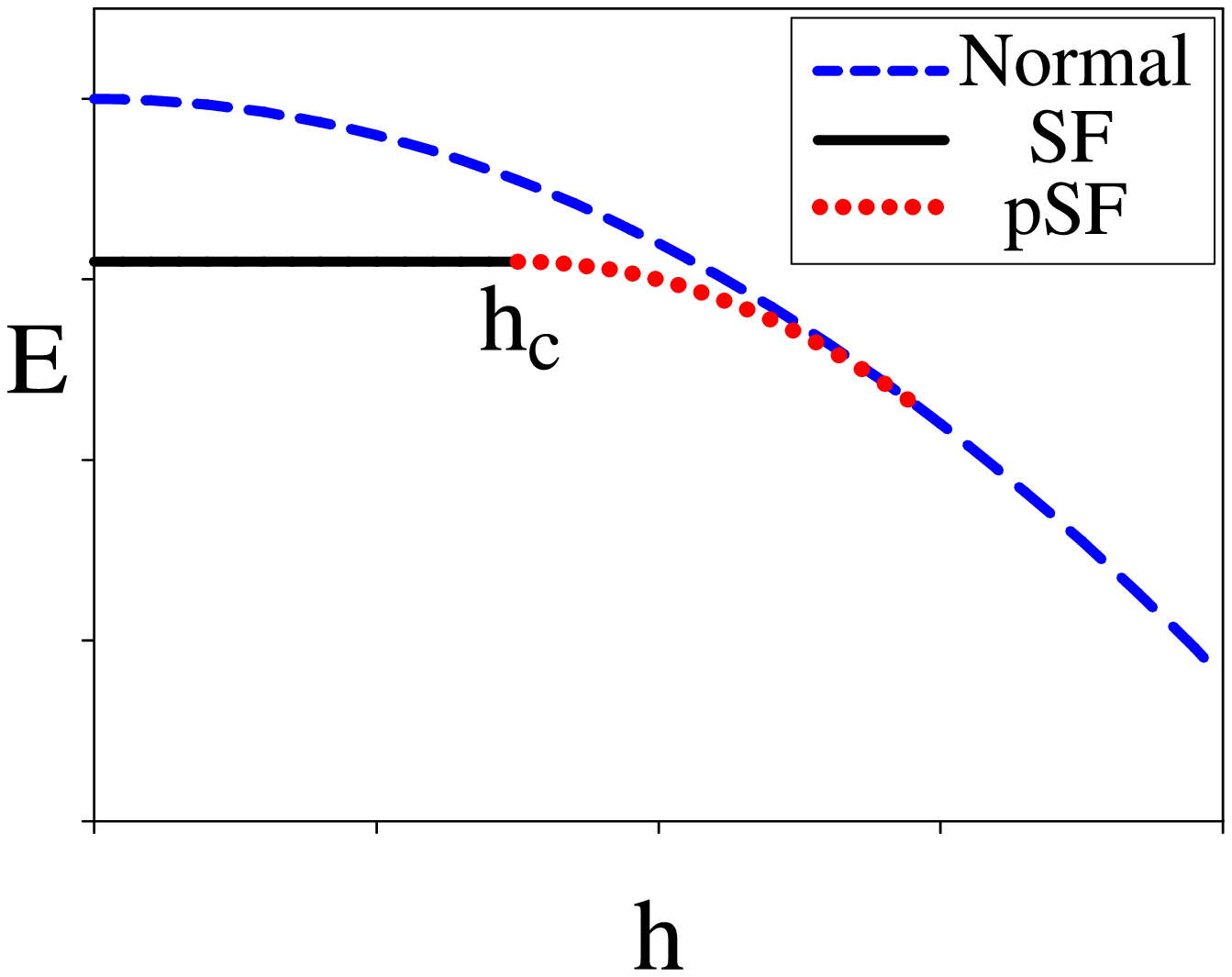}
  \end{center}
  \caption{(Color online) Sketches of the energy $E$ vs $h$, for
the normal state, the unpolarized superfluid (SF) and the pSF phase.
Two situations can appear as function of
the external parameters ({\it e.g.} the interaction strength).
Left panel (a): the pSF branch is unstable and the system undergoes a phase separation.
Right panel (b): the pSF branch is stable.}
  \label{fg:energyconfig}
\end{figure}

We start with some energetic considerations, which clarify the
general conditions under which a  pSF phase can be stable at
$T=0$. In order to control the imbalance between the populations
of the two species, we introduce a chemical potential difference
(or effective `magnetic field') $h\equiv
(\mu_{\uparrow}-\mu_{\downarrow})/2$ between them. In
Fig.~\ref{fg:energyconfig}, we show two typical behaviors of the
energy in different phases as a function of the magnetic field.
In both cases, a small magnetic field $h$ is expelled from the
unpolarized superfluid, and the energy is independent of $h$. This
unpolarized superfluid is locally stable up to a critical value
$h_c$. For $h > h_c$ the magnetic field breaks the pairs, leading
to the disappearance of this solution. On the other hand, the
energy of the polarized normal state is a decreasing function of
$h$: its derivative $p\equiv -\partial E/\partial h >0$ is the
polarization (population imbalance) and its curvature $\chi \equiv
-\partial^2 E / \partial h^2 =
\partial p/\partial h$ defines the polarizability of the normal
fluid.

In general the pSF phase can bridge between these two solutions.
The way in which this connection occurs depends on the two key
parameters $h_c$ and $\chi$. When $h_c$ and $\chi$ are large, we
anticipate the situation in Fig.~\ref{fg:energyconfig}a. In this
case, the pSF branch is expected not to be stable, and the system
undergoes a first-order transition as a function of $h$ which
results in a phase separation if we try to prepare the system with
a polarization corresponding to the unstable branch. In contrast,
if $h_c$ and $\chi$ are small enough, the energies of the
unpolarized superfluid and polarized normal solutions do not
cross, and the pSF phase can be stable in a region bridging these
two states, as shown in Fig.~\ref{fg:energyconfig}b. Therefore, a
stable pSF phase is likely to form when $\chi$ or $h_c$ are small.
Interestingly, this suggests that an increasing attractive
coupling may help stabilizing the pSF phase. Indeed, in the BEC
regime, the normal state presents preformed pairs in a singlet
state that strongly reduce $\chi$, hence stabilizing a pSF phase.

In order to explore the validity of these qualitative arguments
we study an attractive Hubbard model at half-filling,
on a three-dimensional cubic lattice with nearest-neighbor hopping:
\begin{align}
\label{hubbard} {\cal H} &= -t \sum_{<ij>\sigma}
(c_{i\sigma}^{\dagger} c_{j\sigma}^\pdag +\mathrm{h.c.}) -
U\sum_{i}n_{i\uparrow}n_{i\downarrow}-\sum_i
\mu_{\sigma}n_{i\sigma} \nonumber
\end{align}
where $c_{i\sigma}^{\dagger}$ ($c_{i\sigma}^\pdag$) creates
(destroys) a fermion of species $\sigma$ on the site $i$,
$n_{i\sigma} = c_{i\sigma}^{\dagger}c_{i\sigma}^\pdag$ is the
number operator, $t$ is the hopping amplitude and $U>0$ is the
Hubbard on-site attraction. When the total number of fermions is
identical to the number of lattice sites (half-filling)
$\mu_{\uparrow}=-U/2+h$ and $\mu_{\downarrow}=-U/2-h$. In the
following, all energies will be expressed in units of the
half-bandwidth $D=6t=1$. We analyze the model within dynamical
mean-field theory (DMFT)~\cite{georges:rmp:1996:1}, which realizes
a quantum (dynamical) mean field of the lattice model in terms of
a single correlated site embedded in a self-consistent bath. This
correlated local problem is then solved using continuous-time
quantum Monte Carlo (CTQMC)~\cite{werner:prl:2006:1}. Contrary to
static mean-field approximations, whose validity is expected to be
limited to weak interactions, DMFT allows to study all the
interaction regimes~\cite{georges:rmp:1996:1}. We compare the DMFT
results with simpler static mean-field calculations, namely with a
standard BCS mean field and a more accurate `BCS-Stoner' mean
field~\cite{CMora,koponen:njp:2008}, which introduces a mean-field
decoupling of the interaction both in the particle-particle
channel (as in BCS) and in the particle-hole channel (as in Stoner
theory) in order to compute both the superfluid order parameter
and the polarization self-consistently.

\begin{figure}[ht!]
  \begin{center}
    \includegraphics[width=7cm,clip=true]{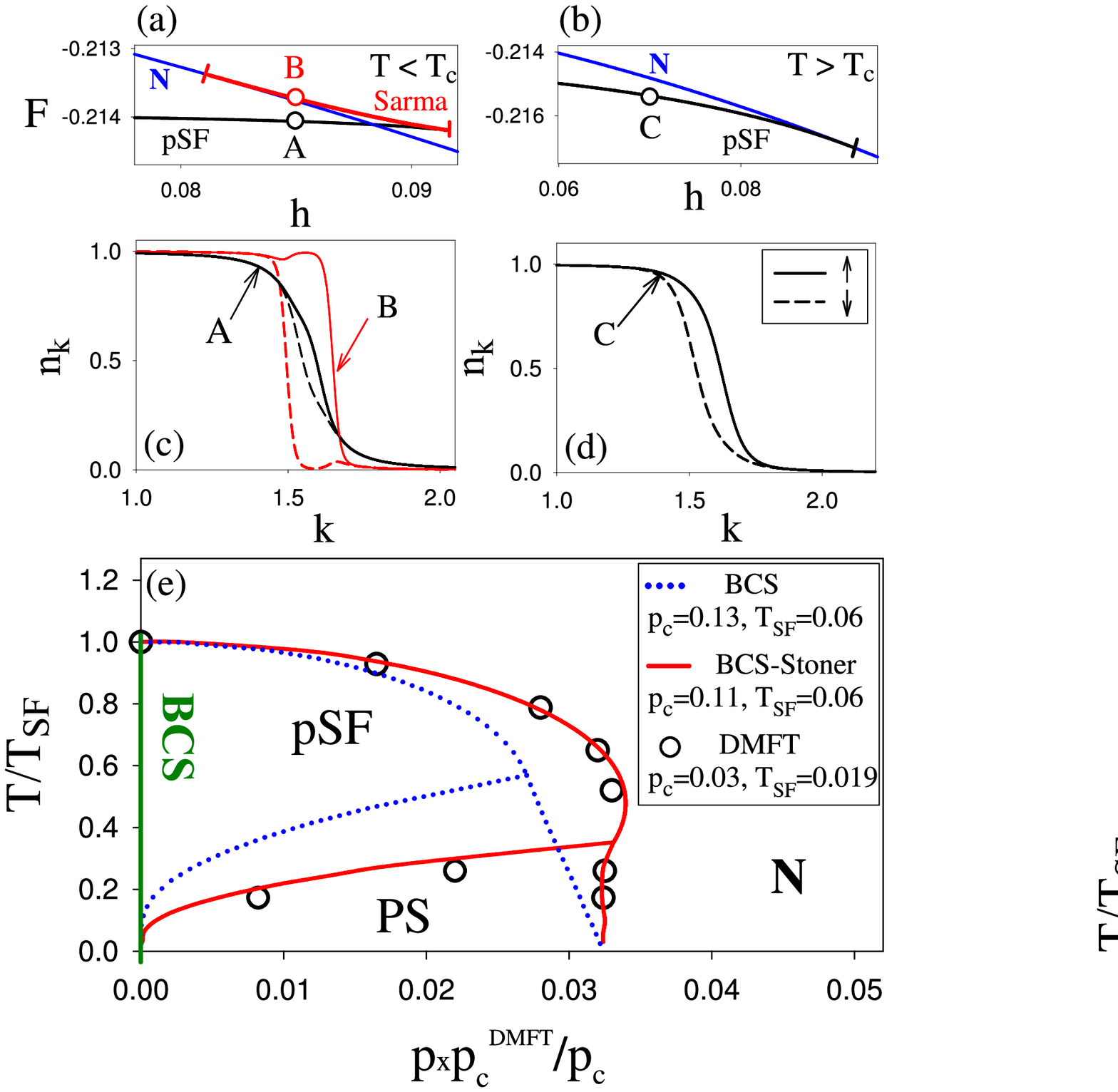}
  \end{center}
  \caption{(Color online) Lower panel (e): Phase diagram in the $T-p$
plane at weak coupling $U = 0.5$ (lower panel) obtained using
DMFT, BCS and BCS-Stoner mean-field. $T_{\SF}$ and $p_{c}$ are
defined in the text. pSF, PS and N label polarized superfluid,
phase separation and normal phase, resp. The results are plotted
against $p/(p_{c} / p_{c}^{\DMFT})$ to allow for a comparison in
relative units. Upper panels (a-b): free-energy {\it vs.} $h$
below and above the critical temperature $T_c$. Middle panels
(c-d): momentum distribution $n(k)$ at three points (A), (B), (C)
indicated on the free-energy curves of panel (a).}
  \label{fg:weakcoupling}
\end{figure}

We first consider a rather weak coupling $U=0.5$. The phase
diagram obtained by using the BCS, BCS-Stoner and DMFT approaches
is presented in Fig.~\ref{fg:weakcoupling}e.
%
%
For large $p$ or at high $T$ the stable phase is the polarized
normal fluid. As $T$ is decreased the system enters a pSF phase
which exhibits both a non-zero superfluid order parameter
$\Delta=(U/N) \langle \sum_i
c^{\dagger}_{i\uparrow}c^{\dagger}_{i\downarrow}\rangle$ and a
finite polarization.  When $T$ is further lowered, the pSF phase
becomes unstable towards a phase separation between a thermally
excited BCS superfluid and a polarized normal fluid. While the
overall phase diagram is the same in all approaches in relative
units (defined below), the BCS mean-field underestimates the
extent of the pSF phase with respects to DMFT, mainly because it
overestimates $\chi$ in the normal state. This effect is
substantially reduced by the BCS-Stoner mean-field, in which the
population imbalance is determined self-consistently. This leads
to a lower $\chi$, which extends the stability of the pSF phase
and improves the agreement with DMFT. Note that, for each
approach, the temperature is normalized by $T_{\SF}$, the
superfluid critical temperature at $h=0$. The polarization is
normalized by $p_c/p_{c}^{\DMFT}$, where $p_{c}$ is the
polarization of the normal phase at $T=0, h= h_{c}$ and
$p_{c}^{\DMFT}$ is the value of $p_{c}$ obtained with DMFT. The
values of $T_{\SF}$ and $p_{c}$ are overestimated in static
mean-field approximations, making a comparison in relative units
more appropriate. In these units, the BCS-Stoner phase diagram is
seen to be in good agreement with the DMFT result in this
weak-coupling regime.

Let us now discuss the nature of these phases. The BCS-Stoner
mean-field calculation shows that in the phase-separated region,
with $T<T_c$, the free-energy as a function of $h$ has three
branches (Fig.~\ref{fg:weakcoupling}a) as in the scenario of
Fig.~\ref{fg:energyconfig}a. If $T$ is small, the properties of
the three branches are directly linked to their $T=0$
counterparts. One branch corresponds to the BCS superfluid with
thermal excitations. It has a small polarization that comes from
thermally excited Bogoliubov quasiparticles in a small
momentum-range around the Fermi momentum $k_F$ of the unpolarized
state.  As a consequence, the density $n(k)$ deviates from the
standard BCS distribution around $k_F$ over a range of order
$T/v_F$ (see A in Fig.~\ref{fg:weakcoupling}c). This branch is
connected to the unstable thermally excited Sarma phase. In
contrast to the BCS state, the Sarma phase has two Fermi surfaces
at $T=0$, which are individually broadened when $T>0$. This is
clearly visible in $n(k)$ (see B in Fig.~\ref{fg:weakcoupling}c)
which displays two humps associated with each Fermi momentum, with
a separation set by the polarization instead of the thermal
broadening.

As the temperature $T$ is increased, the unstable branch becomes smaller and
eventually disappears at $T=T_c$. For $T>T_c$, the pSF phase is stable and the
free-energy has the behavior shown in Fig.~\ref{fg:weakcoupling}b with only two
solutions. Because $T_c$ is rather large, there is no clear distinction between
the thermally excited BCS and the Sarma phases: as $h$ is increased along the superfluid
branch a crossover takes place between the BCS regime and
the Sarma regime. However, because $T$ is large, no
particular structure appears in the density $n(k)$, even close to the normal
phase (see C in Fig.~\ref{fg:weakcoupling}d). Therefore, at weak coupling,
the stable pSF phase has essentially a thermal nature and its properties cannot
be linked to the physics of the Sarma phase.

We now turn to an intermediate coupling $U = 2.5$, where for
identical populations the superfluid state is on the BEC side of
the BCS/BEC
crossover~\cite{toschi:njp,toschi:235118,kyung:024501}. In this
regime, the static mean-field approximations are not expected to
be accurate and we only describe our DMFT results.
%
\begin{figure}[ht!]
  \begin{center}
    \includegraphics[width=7cm,clip=true]{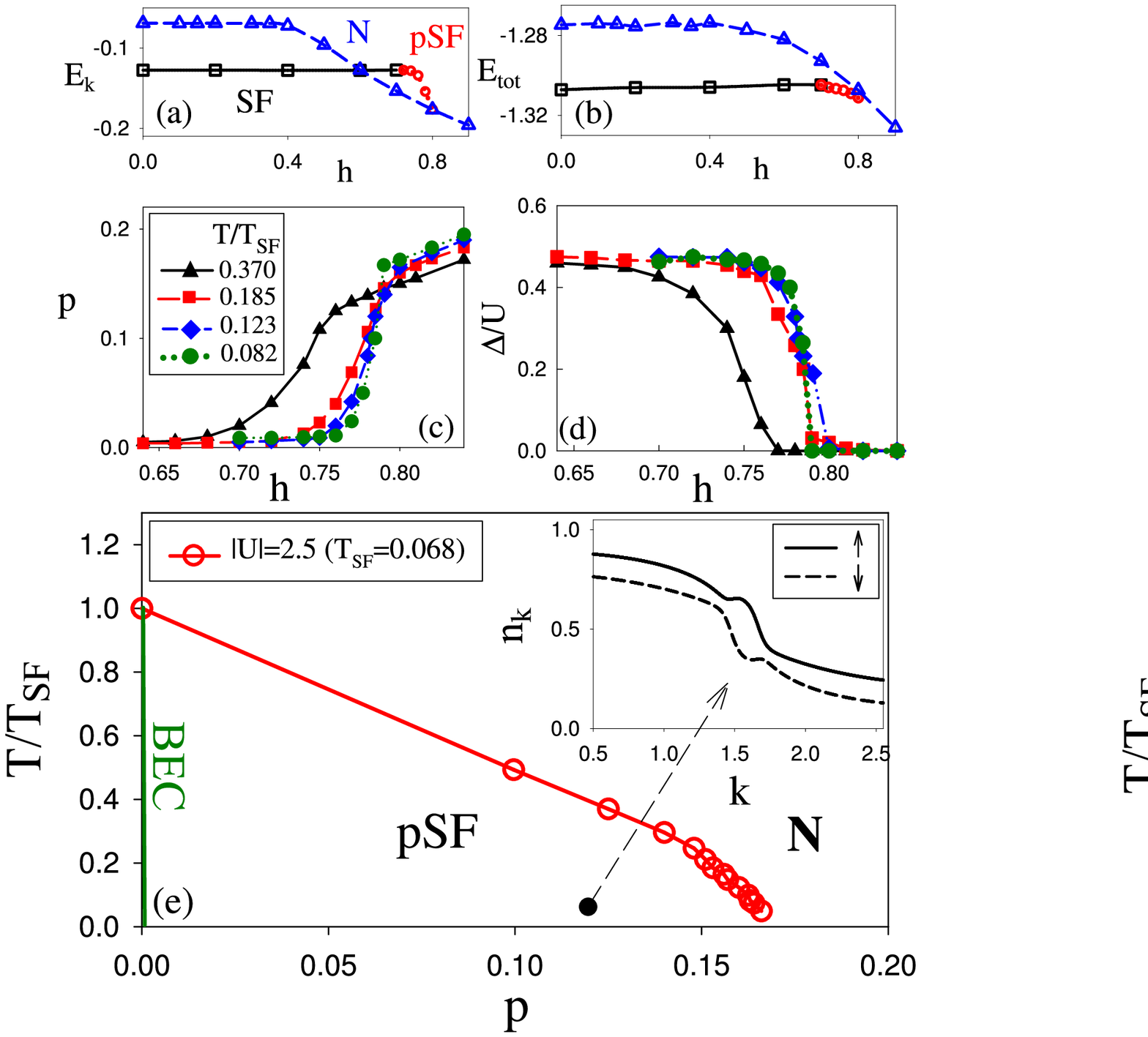}
  \end{center}
  \caption{(Color online) Lower panel (e): Phase diagram in the $T-p$ plane at
intermediate coupling $U= 2.5$ obtained using DMFT.
Inset of (e): momentum distribtion $n(k)$ for $p=0.12$ at the lowest temperature
$T/T_{\SF} = 0.049$.
Upper panels: kinetic energy $E_{k}$ (a) and total internal energy
$E_{tot}$ (b) as a function of $h$ for $T=0.148\ T_{\SF}$.
Middle panels: polarization (c) and superfluid order
parameter (d) as a function of $h$.}
\label{fg:intermediate}
\end{figure}
%
As is clear from Fig.~\ref{fg:intermediate}e, the interaction
strongly increases the stability region of the pSF phase compared to the small
$U$ case: it exists for a larger range of polarization (up to $p \lesssim 16$
\% instead of $p \lesssim 3$\% for $U=0.5$) and is stable  down to the lowest
temperature we could investigate with DMFT ($T/T_{\SF} = 0.049$). From our
present CTQMC solutions of DMFT, we cannot determine whether a phase
separation eventually appears at lower temperatures, as in the weak coupling
regime, but extrapolations of our numerical data are consistent
with a stable pSF phase down to $T=0$.

In Fig.~\ref{fg:intermediate}c-d we plot the superfluid order parameter
$\Delta$ and the polarization as a function of $h$ for different temperatures.
At high temperatures, the polarization gradually increases with $h$ and the pSF
phase smoothly connects to the normal phase. As the temperature is decreased,
two regimes appear in the pSF phase, even though there is no phase transition
between them.  At small $h \lesssim 0.75$, the polarization is very small and
can be traced back to thermal excitations in the BEC state. Around $h \sim
0.75$ a {\it stable branch} connects to the normal phase. The polarization in
this branch is too large to originate from thermal fluctuations and it has a
different nature.  Indeed, the density $n(k)$ in this region (Inset in
Fig.~\ref{fg:intermediate}e) displays two humps, just like in the weak-coupling
Sarma phase (Fig.~\ref{fg:weakcoupling}). This is very different from what
is expected at low temperature in a standard thermally excited superfluid
where $n(k)$ is broadened around $k_F$ over a small range $\sim T/v_F$. The two
humps also indicate that the underlying $T=0$ phase has two Fermi
surfaces, unlike the BP1 phase proposed deep in the BEC regime of trapped
fermionic gases. This shows that at intermediate couplings on a lattice, it is
not possible to reduce the problem to a simple Bose-Fermi mixture.

Hence, our results show that a larger coupling stabilizes a region
which displays properties very similar to the Sarma phase
discussed at weak coupling, in agreement with the qualitative
energetic arguments that a reduced polarizability and preformed
pairs help stabilizing the pSF phase at low temperatures. This is
actually confirmed by a direct computation of the energetic
balance underlying this stabilization. The total internal energy
and the kinetic energy of each phase are displayed in
Fig.~\ref{fg:intermediate}a-b as a function of $h$, for $U = 2.5$
and $T=0.148\ T_{\SF}$. For this very low temperature, the entropy
term can be neglected and we consider the energy instead of the
free-energy. The total energy curve nicely follows the second
scenario described above (Fig.~\ref{fg:energyconfig}b). A stable
pSF phase bridges between the flat energy of the unpolarized
superfluid and the energy curve of the polarized normal fluid. The
total energy branch corresponding to the normal phase is seen to
have a reduced curvature in comparison to weaker couplings,
indicating a small $\chi$ of the normal fluid (within DMFT, this
branch has actually vanishing polarization up to to a field $h
\sim 0.4$). These effects strongly favor the stability of the pSF
phase.

The energetic balance of the transition to the pSF state is particularly interesting.
In the absence of imbalance it has been shown that for $U=2.5$
the system is in the BEC regime and the superfluid state is stabilized by a gain of kinetic
energy~\cite{toschi:njp,toschi:235118,kyung:024501}, in contrast with the
BCS state which gains potential energy.
Here we find, as shown in Fig.~\ref{fg:intermediate}a, that the pSF has instead
higher kinetic energy than the normal state and it is therefore stabilized by potential energy,
even though we are not in the BCS regime.
Therefore, as a function of the imbalance of populations, the system will turn
from a regular BEC system which gains kinetic energy in the superfluid state to
a pSF phase which loses kinetic energy.  Measurements of energies are
experimentally possible in cold atomic systems~\cite{stewart:220406} and it
would be of great interest to investigate these energetic considerations for
polarized gases.

In conclusion, general arguments based on energy considerations suggest that a
polarized superfluid phase can be stabilized by the formation of preformed
pairs with a reduced polarizability on the BEC side of the BCS-BEC crossover.
We have substantiated these arguments with a DMFT solution of the half-filled
attractive Hubbard model on the cubic lattice, which demonstrates the stabilization of a
pSF phase down to very low temperatures for an intermediate coupling
$U/(6t)=2.5$. The nature of this phase is closely connected to the physics of
the Sarma (BP2) phase that has been previously discussed at weak coupling
by static mean-field theory, but is usually unstable in this regime.
We have shown that the stabilized pSF phase is clearly distinct from a BP1 phase and from a
standard thermally excited superfluid state.  Finally, while the BEC superfluid
(in contrast to the weak-coupling BCS one) is stabilized by a gain in kinetic energy,
the pSF-phase condensation energy corresponds to a potential energy gain in
comparison to the polarized normal fluid.

\acknowledgments
We thank P. S. Cornaglia, C. Mora, T.-S. Dam and W. V. Liu for useful discussions,
and the Aspen Center for Physics for hospitality.
We acknowledge the support of the Agence Nationale de la Recherche (ANR) under contracts
GASCOR and FABIOLA, the DARPA-OLE program and Italian MIUR PRIN 2007.

\end{document}